\documentstyle[12pt,aps,prc,epsfig,preprint]{revtex}
\tightenlines
\begin{document}

\begin{titlepage}
\title{\vspace*{5mm}\bf
\Large Testing Chiral Dynamics in Pionic Atoms}
\vspace{4pt}

\author{E.~Friedman and A.~Gal \\
{\it Racah Institute of Physics, The Hebrew University,
Jerusalem 91904, Israel\\}}

\vspace{4pt}
\maketitle

\begin{abstract}

The energy dependence of chirally expanded $\pi N$ isoscalar and
isovector amplitudes $b_{0}(E)$ and $b_{1}(E)$ respectively,
for zero-momentum {\it off shell} pions near threshold,
is used to impose the minimal substitution requirement
$E \to E - V_{c}$ on the properly constructed pion optical potential
within a large-scale fit to 100 pionic-atom data
across the periodic table which also include the recently established
`deeply bound' pionic atoms of Pb and Sn. This fit cannot be reconciled
with the well known free-space values of the $\pi N$ threshold amplitudes.
In contrast, introducing the empirically known energy dependence for
{\it on-shell} pions leads to a better fit and to satisfactory
values for the $\pi N$ threshold amplitudes.
The difference between these two approaches is briefly discussed.
\newline
$PACS$: 12.39.Fe; 13.75.Gx; 21.30.Fe; 36.10.Gv
\newline
\newline \newline
Corresponding author: Avraham Gal, \newline
Tel: +972 2 658 4930, FAX: +972 2 5611519, \newline
E mail: avragal@vms.huji.ac.il

\end{abstract}

\centerline{\today}
\end{titlepage}

\section{Introduction and Methodology}
\label{sec:int}

The recent observation of $1s$ and $2p$ `deeply bound' $\pi ^{-}$
atomic states in isotopes of Pb \cite{YHI96,GGK00,GGG02} and very
recently also of such $1s$ states in isotopes of Sn \cite{SFG02}
has triggered renewed interest in the issue of partial restoration
of chiral symmetry in dense nuclear matter
\cite{Wei00,KYa01,Fri02,GGG02a,GNO02,FGa02,FGa03,KKW03,KKW03a}.
In a nutshell, it was argued that since (i) the pion in
deeply bound states with relatively large neutron excess
charts a fairly dense portion of the nuclear medium,
and since (ii) the most influential term of the optical potential
$V_{\rm opt}$ for this situation is generated by the $s$-wave
isovector $\pi N$ threshold amplitude $b_{1}$, and since
(iii) $b_{1}$ in free-space is well approximated in lowest
chiral-expansion order by the Tomozawa-Weinberg expression
\cite{Tom66}

\begin{equation}
\label{equ:b1}
b_{1}=-\frac{\mu_{\pi N}}{8 \pi f^{2}_{\pi}}=-0.08~m^{-1}_{\pi} \,,
\end{equation}
then deeply-bound states could yield valuable information on the
dependence of $f_{\pi}$ on the density $\rho$. The pion decay
constant $f_{\pi}$ serves as an order parameter for the spontaneously
broken chiral symmetry in hadronic physics, and its free-space value
$f_{\pi}$= 92.4 MeV should go to zero in dense matter if and when
chiral symmetry is restored. Indeed, it has been known for quite some
time that the renormalized value of $b_{1}$ required to fit
pionic-atom data is about $-0.12~m^{-1}_{\pi}$ \cite{BBF78,GNO92}
clearly more repulsive than the free-space value $-0.09~m^{-1}_{\pi}$
\cite{SBG01}.

Our most extensive recent work \cite{FGa03} has shown, however,
that the deeply bound states by themselves on statistical grounds
are insufficient to draw firm conclusions about whether or not $b_{1}$
is renormalized in dense matter. In fact, contrary to the expectation
(i) above, the pion in deeply bound $1s$ states {\it does not}
chart higher-density regions of the nucleus than it does so in `normal'
$1s$ states in light nuclei. It was shown in Ref. \cite{FGa03} that
only by using a substantially larger data base that includes plenty
of normal pionic atom data, and carefully considering uncertainties
in the knowledge of neutron density distributions, it becomes possible
to make a meaningful statement on the renormalization of the isovector
threshold amplitude $b_{1}$, i.e. $b_{1}=-0.108\pm 0.007~m_{\pi}^{-1}$.
It is tempting to ascribe this value for $b_{1}$, using
Eq. (\ref{equ:b1}), to a renormalization of $f_{\pi}$
in the nuclear medium.

Recently, Kolomeitsev, Kaiser and Weise \cite{KKW03} have suggested
that pionic atom data could be reproduced using a pion optical
potential underlain by chirally expanded $\pi N$ amplitudes,
retaining the energy dependence of the amplitudes $b_{0}(E)$ and
$b_{1}(E)$ for zero-momentum ({\bf q}=0) pions in nuclear matter
in order to impose the minimal substitution requirement
$E \to E - V_{c}$, where $V_c$ is the Coulomb potential.
This has the advantage of enabling one to use
a systematic chiral expansion as an input \cite{KWe01}, rather than
singling out the leading order term Eq. (\ref{equ:b1}) for $b_{1}$.
Kolomeitsev et al. applied this programme to study the shifts
and widths of the Pb and Sn pionic deeply bound states
and reported substantial improvement in reproducing these data
with only minimal phenomenological input, mostly limited to the
$p$-wave $\pi N$ interaction to which allegedly these data are
insensitive \cite{KKW03,KKW03a}.
One could argue, however, that the deeply bound states do not
offer sufficient variation over the energy range spanned by the
bulk data on pionic atoms which include both `shallow' states
as well as `deep' states.
In the present work we test whether or not the energy dependence
of this chiral expansion provides a satisfactory
description for the bulk of pionic atom data.

The data used in the present work consist of 100 strong-interaction
shifts and widths, stretching from $^{20}$Ne to $^{238}$U \cite{FGa03}.
The Klein-Gordon equation solved for the pionic-atom eigen energies
is given by \cite{FGa03,BFG97}:

\begin{equation}
\label{equ:KG1}
\left[ \nabla^2  - 2{\mu}(B + V_c) + (B + V_c)^2 - \Pi(E)\right] \psi
= 0~~ ~~ ~~ ~~(\hbar = c = 1)
\end{equation}
where $\mu$ is the pion-nucleus reduced mass,
$B$ is the complex binding energy and $V_c$ is the finite-size
Coulomb interaction of the pion with the nucleus, including
vacuum-polarization terms.
The pion-nuclear polarization operator $\Pi(E)$
is given by the standard Ericson-Ericson form \cite{EEr66}

\begin{equation}
\label{equ:EE1}
\Pi = 2{\mu}V_{\rm opt} = q(r) + \vec \nabla \cdot \alpha(r) \vec \nabla \,,
\end{equation}
with the $s$-wave part of $V_{\rm opt}$
\begin{eqnarray}
\label{equ:EE1s}
q(r) & = & -4\pi(1+\frac{\mu}{M})\{{\bar b_0}(r)[\rho_n(r)+\rho_p(r)]
 +b_1[\rho_n(r)-\rho_p(r)] \} \nonumber \\
 & &  -4\pi(1+\frac{\mu}{2M})4B_0\rho_n(r) \rho_p(r) \,.
\end{eqnarray}
In these expressions $\rho_n$ and $\rho_p$ are the neutron and proton
density distributions normalized to the number of neutrons $N$ and
number of protons $Z$, respectively, and $M$ is the mass of the nucleon;
$q(r)$ is referred to as the $s$-wave potential term and $\alpha(r)$ is
referred to as the $p$-wave potential term.
The function ${\bar b_0}(r)$ in Eq. (\ref{equ:EE1s})
is given in terms of the {\it local} Fermi
momentum $k_{\rm F}(r)$ corresponding to the isoscalar nucleon
density distribution:

\begin{equation}
\label{equ:b0b}
{\bar b_0}(r) = b_0 - \frac{3}{2\pi}(b_0^2+2b_1^2)k_{\rm F}(r) \,,
\end{equation}
where the quadratic terms in $b_0$ and $b_1$ represent double-scattering
modifications of $b_0$. In particular, the $b_1^2$ term represents
a sizable correction to the nearly vanishing linear $b_0$ term.
Similar double-scattering modifications of $b_1$, as well as other
correction terms to $\Pi (E)$ listed in Refs. \cite{KKW03,KWe01},
were found by us to yield negligibly small effects and will not be
further discussed below. The complex parameter $B_0$ is due to $s$-wave
absorption on pairs of nucleons.
Its microscopic evaluation is outside the scope of chiral perturbation
theory. Finally, the $p$-wave term $\alpha(r)$ is a standard one with
the same form as in Ref. \cite{FGa03}.

The chiral expansion of the $\pi N$ amplitudes for {\bf q} = 0 at the
two-loop level is well approximated by the following expressions
\cite{KKW03,KWe01}:
\begin{equation}
4\pi \left( {1+{{m_\pi } \over M}} \right)b_{0}(E)\approx
 \left( {{{\sigma -\beta E^2} \over {f_\pi ^2}}+{{3g_A^2m_\pi ^3} \over
{16\pi f_\pi ^4}}} \right)\   ,
	\label{eqn:b0ch}
\end{equation}

\begin{equation}
4\pi \left( {1+{{m_\pi } \over M}} \right)b_{1}(E)\approx
-{E \over {2f_\pi ^2}} \left( {1+{{\gamma E^2} \over {\left( {2\pi f_\pi }
\right)^2}}} \right)\   ,
	\label{eqn:b1ch}
\end{equation}
where $\sigma$ is the $\pi N$ sigma term, $\sigma \sim 50$ MeV
\cite{San02}, $g_{A}$ is the nucleon axial-vector coupling constant,
$g_{A}=1.27$, $\beta$ and $\gamma$ are tuned to reproduce the
threshold values $b_{0}(m_{\pi}) \approx 0$ and $b_{1}(m_{\pi}) =
-0.0885^{+0.0010}_{-0.0021}~~m^{-1}_{\pi}$ \cite{SBG01} respectively.
For $b_{0}$, in view of the accidental cancellations that lead to
its near vanishing we limit our discussion to the $f_{\pi}^{-2}$
term in Eq. (\ref{eqn:b0ch}), therefore choosing
$\beta = \sigma m_{\pi}^{-2}$. The next, $f_{\pi}^{-4}$ term
is much bigger than the scale of variation from zero expected for
the threshold value and its inclusion here would appear somewhat dubious;
if included, it would increase the energy dependence from the
conservative estimate adopted by us.
Implementing the minimal substitution requirement in
the calculation of pionic atom observables, the constant parameters
$b_{0,1}$ of the conventionally energy independent optical potential
have been replaced in our calculation by

\begin{equation}
\label{equ:b01}
b_{0,1}(r)=b_{0,1} - \delta _{0,1} ({\rm Re}B + V_c(r)) \,,
\end{equation}
where $\delta _{0,1} = \partial b_{0,1}(E)/\partial E$ is the
appropriate slope parameter at threshold, Re$B$ is the (real)
binding energy of the corresponding pionic atom state and $V_c(r)$
is the Coulomb potential. The constant fit parameters $b_{0,1}$ are
then expected to agree with the corresponding free $\pi N$ threshold
amplitudes if the energy dependence is indeed responsible for the
renormalized values found in conventional analyses.
The added piece proportional to $\delta$ in Eq. (\ref{equ:b01}) is
dominated by the attractive $V_c(r)$. Since the slope parameters
$\delta$ from Eqs. (\ref{eqn:b0ch},\ref{eqn:b1ch}) are negative,
this added piece is always repulsive, in agreement with
Refs. \cite{KKW03,ETa82}.

\begin{figure}
\epsfig{file=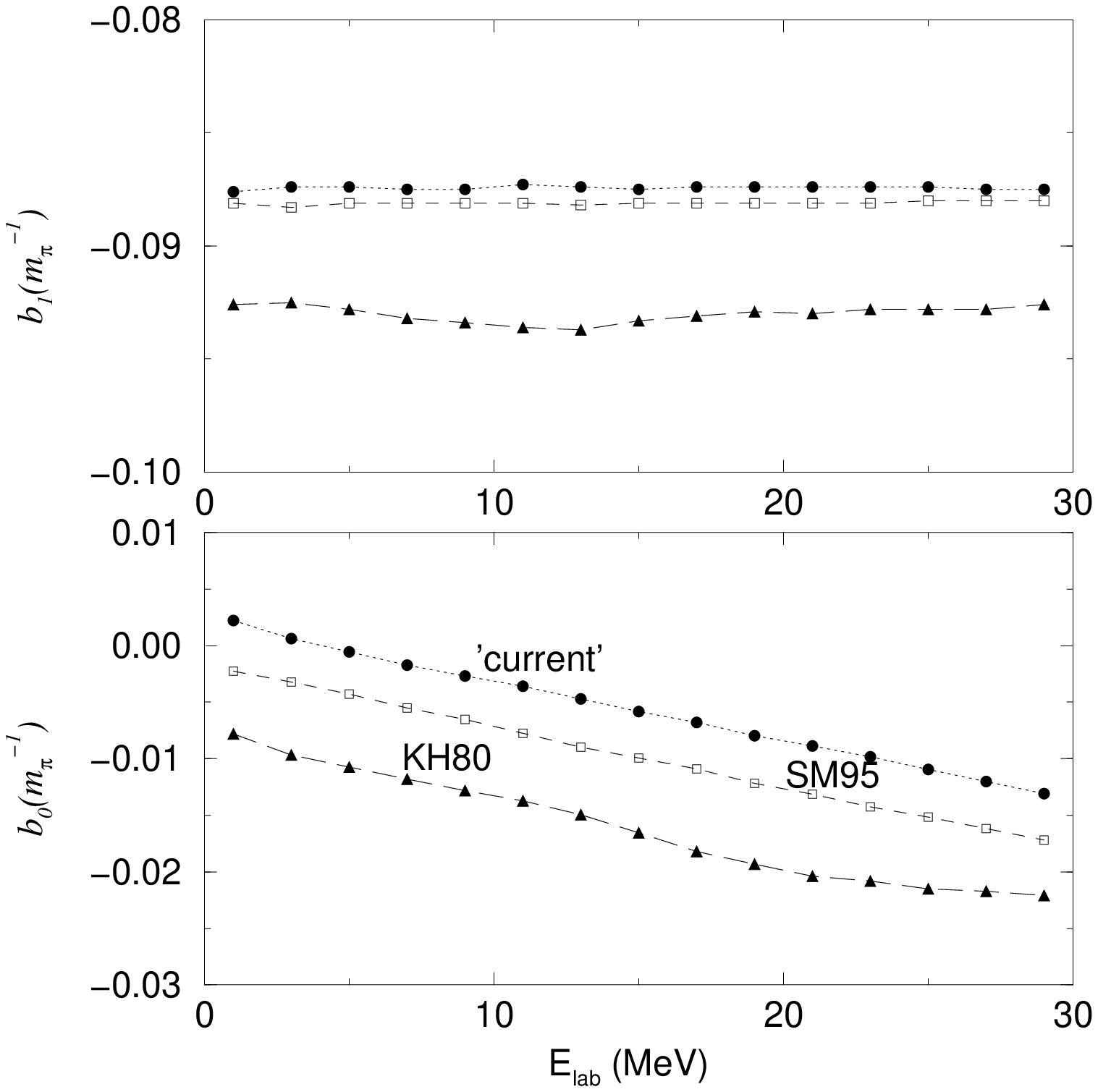, height=140mm,width=150mm}
\caption{$\pi N$ empirical $s$-wave scattering amplitudes as function of
laboratory energy for on-shell pions from the SAID data base
{\protect \cite{SAID}}.}
\label{fig:phases}
\end{figure}

Before testing the above `chiral' energy dependence for {\it off-shell}
{\bf q} = 0 pions we present results for the empirically known
{\it on-shell} $\pi N$ amplitudes, when the pion energy $E$ and its
three-momentum {\bf q} are related by $E^2 = m_{\pi}^2 + {\bf q}^2$.
This choice corresponds to the original suggestion by Ericson and
Tauscher \cite{ETa82} to consider the effect of energy dependence
in pionic atoms. Ericson subsequently \cite{Eri94} pointed out that,
for strongly repulsive short-range $NN$ correlations, the on-shell
requirement follows naturally from the Agassi-Gal theorem \cite{AGa73}
for scattering off non-overlapping nucleons. The corresponding
$\pi N$ amplitudes will be denoted below as `empirical'.
Figure \ref{fig:phases} shows the energy dependence of the empirical
$b_{0}(E)$ and $b_{1}(E)$ on-shell amplitudes as derived from the
SAID data base \cite{SAID}.
The value of $b_{0}(E)$ at threshold is very close to zero
and the empirical slope $\delta_0$, which corresponds to adding repulsion
in Eq. (\ref{equ:b01}), is quite well-determined over the whole relevant
energy range, having changed little since the classical KH80 analysis
\cite{KPi80} of the pre pion-factories data to the most recent analyses
of modern data. We note that the slope of the {\bf q} = 0 chiral
$b_{0}(E)$ amplitude of Eq. (\ref{eqn:b0ch}) is larger than the on-shell
empirical slope by about 60\%. For the empirical $b_{1}(E)$, its value
at threshold has also changed little since KH80 to the present day
analysis, and the slope of the empirical amplitude is essentially zero,
in contrast to the fairly large slope of the {\bf q} = 0 chiral amplitude
of Eq. (\ref{eqn:b1ch}).

\section{Results}
\label{sec:res}

The present analysis is based on the `global 3' data set of
Ref. \cite{FGa03} consisting of 100 data points from $^{20}$Ne to
$^{238}$U. For the nuclear density distributions $\rho _p$ and $\rho _n$
we adopt the procedure of Ref. \cite{FGa03} where $\rho _p$ is obtained
from the experimental charge distribution by unfolding the finite size
of the charge of the proton, and where simple but physical
parameterizations are used for $\rho _n$. A key quantity in this
context is the difference $r_n-r_p$ between the root-mean-square radii.
Relativistic mean field (RMF) calculations \cite{LRR99} yield to
a good approximation \cite{FGa03}

\begin{equation}
\label{equ:RMF}
r_n-r_p = \alpha \frac{N-Z}{A} + \eta \,,
\end{equation}
with the values $\alpha$=1.51$\pm$0.07~fm, $\eta$=$-$0.03$\pm$0.01 fm.
A similar expression, but with $\alpha=1.0$, was obtained by analyzing
strong interaction effects in antiprotonic atoms \cite{TJL01}.
Owing to the strong correlation between the values assumed for $r_n-r_p$
and the values of $b_1$ derived from $\chi ^2$ fits to pionic atom  data,
we have varied the neutron-excess parameter $\alpha$ over a wide range,
with the expectation that a value in the range of 1.0 to 1.5 will
represent {\it on the average} the 41 nuclei in the present data base.

\begin{figure}
\epsfig{file=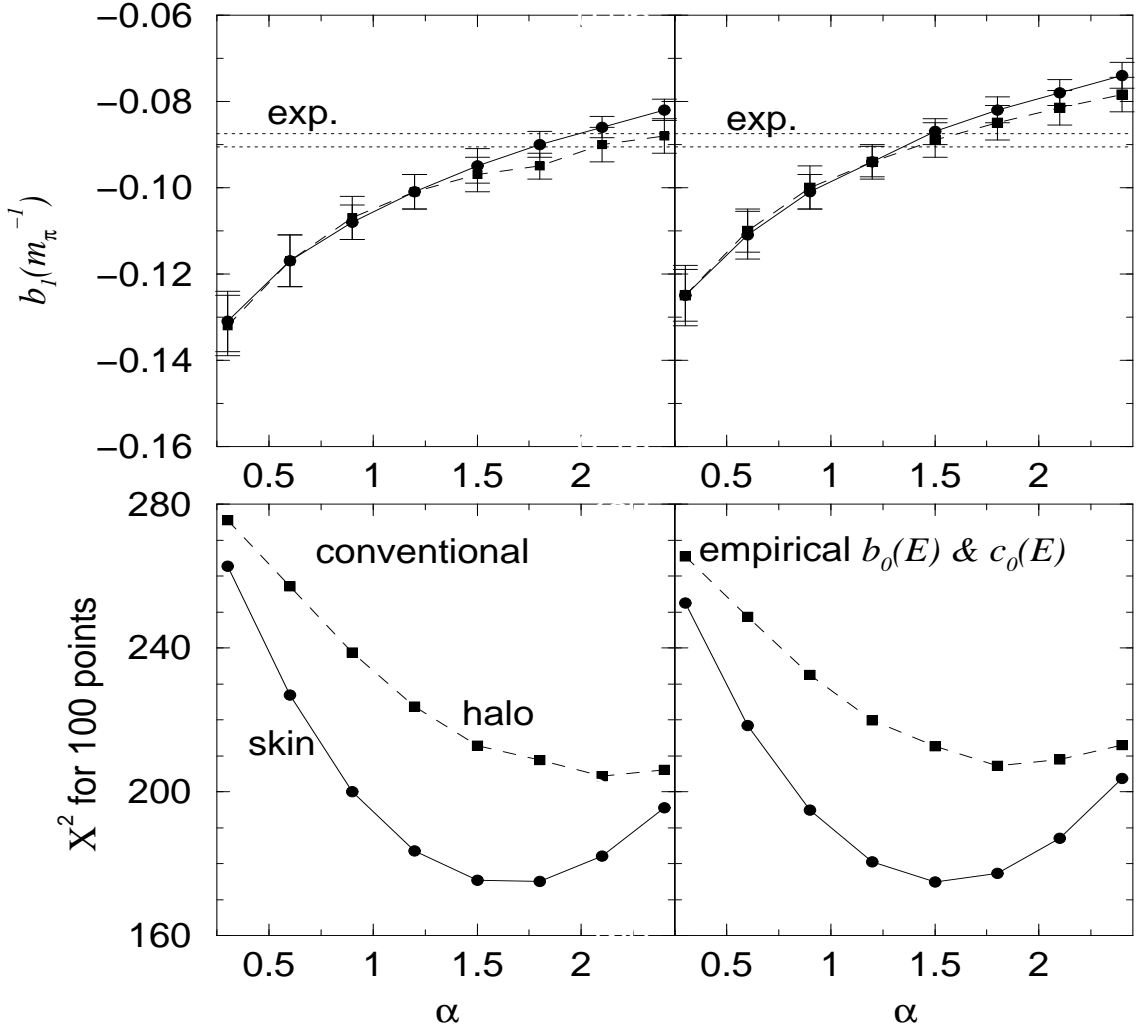, height=140mm,width=150mm}
\caption{Pionic-atom fits using `conventional', energy independent
$\pi N$ amplitudes (left panels) and `empirical', energy dependent
SAID amplitudes (right panels) as function of the
neutron-excess parameter $\alpha $, Eq.(\ref{equ:RMF}),
for two shapes of neutron densities. Lower part: values of $\chi ^2$
for 100 data points from $^{20}$Ne to $^{238}$U. Upper part:
best-fit values of $b_1$ vs. the free $\pi N$ threshold value
{\protect \cite{SBG01}} marked by `exp.' within the dotted horizontal
lines.}
\label{fig:convemp}
\end{figure}

Figure \ref{fig:convemp} shows results for the `conventional' model
(left) for which $\delta_{0,1}=0$ and for the `empirical' model (right)
as function of the neutron-excess parameter $\alpha$ in
Eq. (\ref{equ:RMF}) for two shapes of neutron densities.
The dependence of the quality of fits on the shape of the neutron density
distribution and the vanishingly small sensitivity of the derived values
of $b_1$ to this shape are demonstrated by using either the `skin' or
the `halo' shape, as discussed in Ref. \cite{FGa03}.
The results for the conventional (energy independent) model
are practically the same as in Ref. \cite{FGa03} in spite of adopting
now the `current' SAID values \cite{SAID} for the $p$-wave
parameters $c_0 = 0.21~~m_\pi ^{-3}$ and $c_1 = 0.165~~m_\pi ^{-3}$,
instead of the values 0.22 and 0.18 $m_\pi ^{-3}$ respectively
used beforehand. This slight change was made for consistency,
since the slope parameters $\delta $ for the empirical model were
taken from the `current' SAID analysis. In fact, we also incorporated in
the right hand side (rhs) of the figure the SAID weak energy dependence of
$c_0$ ($c_1$ is essentially energy independent). The results on the rhs
of the figure show that, for the `skin' shape of
$\rho _n$ and with the introduction of the empirical energy dependence
of the amplitudes, the minimum in the $\chi ^2$ curve has shifted
slightly towards the acceptable region of $\alpha=$1.0 to 1.5, and the
value of $b_1$ for that minimum is in agreement with the free $\pi N$
value marked by `exp.' in the upper panels of the figure.
It is self-evident that the `halo' shape for the neutron density
distributions cannot be reconciled with the data. Finally, we add that
the resulting values of $b_0$ for both the conventional and empirical
models are close to zero, well within the experimental error \cite{SBG01}.

A comment on the `anomalous $s$-wave repulsion' in pionic atoms is here
in order. The net effect of the nearly vanishing parameter $b_0$,
of the repulsive $b_1$
and of the phenomenological parameter Re$B_0$ has been known \cite{BFG97}
to produce a repulsive potential inside nuclei which is twice as large
as expected. This is due to the combined action of the too repulsive
$b_1$ and of Re$B_0$ which
turns out to be too repulsive compared to the expectations that
$|$Re$B_0| < $Im$B_0$ (see also Ref. \cite{Fri02}).
For the fits mentioned above we obtain for the conventional potential
Im$B_0$=0.053$\pm$0.002 $m^{-4}_{\pi}$
and Re$B_0=-$0.10 $\pm$ 0.03 $m^{-4}_{\pi}$.
Although the latter is determined to a moderate accuracy,
we note that setting its value to zero while repeating the fits leads to a
significant increase in the resulting $\chi ^2$ value and to a value for
$b_0$ which is incompatible with experiment (cf. Table 4 of
Ref. \cite{FGa03}). Using the empirical $b_{0,1}(E)$ we find
Re$B_0=-$0.07 $\pm$ 0.03 $m^{-4}_{\pi}$. We conclude that using
the empirical energy dependence the anomaly in Re$B_0$ is reduced,
whereas there is essentially no anomaly in the parameter $b_1$.

\begin{figure}
\epsfig{file=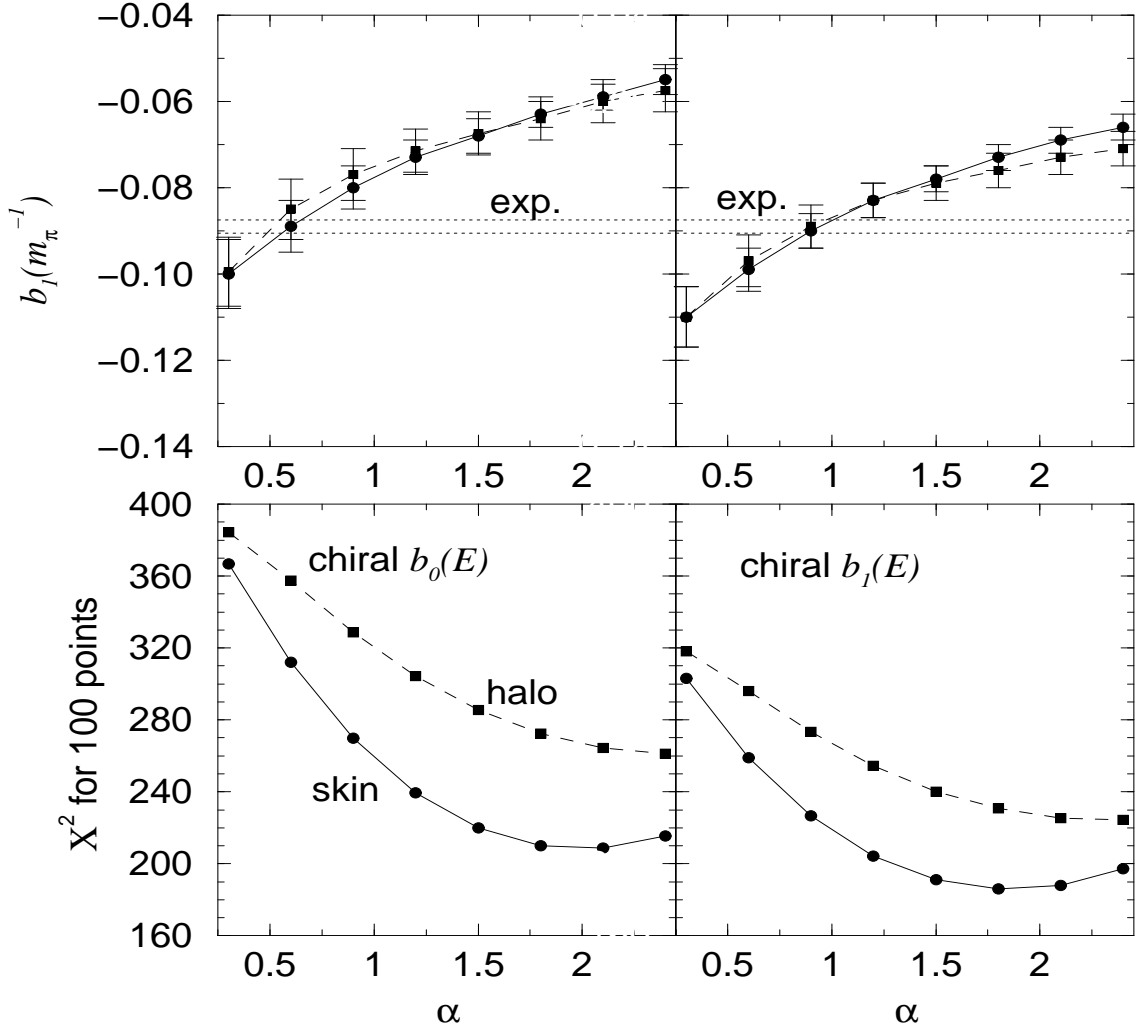, height=140mm,width=150mm}
\caption{Same as Fig. \ref{fig:convemp} but for the energy dependent
`chiral' amplitudes $b_0$ (left) and $b_1$ (right).}
\label{fig:chiral}
\end{figure}

Figure \ref{fig:chiral} shows results for the `chiral' model,
when either $b_0$ (left) or $b_1$ (right) is made energy dependent
according to Eqs. (\ref {eqn:b0ch}) and (\ref {eqn:b1ch}), respectively.
The left hand side of the figure shows that the quality of the best fit,
upon incorporating only the energy dependence of the chiral $b_0$ amplitude,
is significantly inferior to the corresponding best fit obtained using the
conventional, energy independent model (shown in Fig. \ref{fig:convemp}).
Furthermore, the value of $\alpha$ at the $\chi ^2$ minimum is unacceptably
large and the corresponding value of $b_1$ is in sharp disagreement with
the experimental free $\pi N$ threshold value. The rhs of the figure shows
good fits with almost acceptable values for $\alpha$ and for $b_1$, upon
incorporating only the energy dependence of the chiral $b_1$ amplitude,
again for the `skin' shape of the neutron density.
However, incorporating the energy dependence of the chiral
$b_{0}(E)$ (even within the limited scope of using only the $f_{\pi}^{-2}$
term on the rhs of Eq. (\ref{eqn:b0ch}))
on top of that for $b_{1}(E)$, leads to substantial disagreement
between the resulting best-fit value for $b_1$ and the threshold value
$b_{1}(m_{\pi})$ which is marked by `exp.' in Fig. \ref{fig:chiral}.
We conclude that, at present, the energy dependence generated by the
chirally expanded $s$-wave $\pi N$ amplitudes of
Eqs. (\ref{eqn:b0ch},\ref{eqn:b1ch}) fails badly in reproducing
consistently the bulk of pionic atom data.

\section{Discussion and Conclusions}
\label{sec:dis}

In the present work we have demonstrated
that the consistency between pionic-atom data and the free $\pi N$
threshold amplitudes is greatly improved by using just
the on-shell energy dependence of the $\pi N$ $s$-wave amplitudes,
in accordance with the original suggestion made by Ericson and Tauscher
\cite{ETa82}. The idea behind using this empirical energy dependence
is the same one as used for constructing the multiple-scattering series
for short-ranged $\pi N$ interactions occurring within an assembly of
largely non-overlapping nucleons \cite{EEr66,Eri94}.
Multiple scattering is naturally described in this idealized limit
as occurring {\it on shell}. Whereas using off-shell {\bf q}=0 pions
in chiral expansions is motivated by the ground-state wavefunction
description of pions in {\it nuclear matter}, applying this limitation
to the construction of the pion-nuclear optical potential that generates
pionic-atom wavefunctions is questionable.

We have also shown that the energy-dependent chiral amplitudes given by
Eqs. (\ref {eqn:b0ch},\ref {eqn:b1ch}) for {\bf q}=0 off-shell pions
do not produce consistent or good global fits to pionic atom data.
This conclusion is not at odds with the observation made by Kolomeitsev
et al. \cite{KKW03} that the {\bf q}=0 off-shell chiral amplitudes work well,
and with no need for a dispersive term Re$B_0$ for the few deeply bound
states available at present, since partial data sets of this kind do not
have sufficient statistical significance to decide one way or another on
this issue \cite{FGa02,FGa03}. In fact, as good average reproduction of
these deeply-bound data is reached within a wide classs of optical potentials,
including our `empirical' energy-dependent potential of the present study.
We defer this and other ramifications of the present analysis for
a forthcoming detailed publication. Given the fact that the on-shell
$\pi N$ amplitudes provide by far a better description of pionic atom
data than the extremely off-shell {\bf q}=0 chiral amplitudes do,
we conclude that chiral dynamics is not yet at a stage of being tested
in pionic atoms.

Finally, it should be emphasized that we have strictly adhered in the
present calculation to imposing minimal substitution, $E \to E - V_{c}$,
on the pion-nuclear polarization operator $\Pi(E)$ within the Klein-Gordon
equation (\ref{equ:KG1}).
Nowhere have we renormalized the threshold value of the $\pi N$ isovector
amplitude $b_{1}$ of Eq. (\ref{equ:b1}) by renormalizing the pion decay
constant $f_{\pi} \to f_{\pi}(\rho)$ in dense matter \cite {Wei00}. This
latter prescription which appears to be rooted in the underlying chiral
symmetry has been discussed extensively in the context of pionic atoms
\cite{KYa01,Fri02,GGG02a,GNO02,KKW03,KKW03a}, but according to
Refs. \cite{KKW03,KKW03a} it need not be applied once the full energy
dependence of $\Pi(E)$ is incorporated.

\vspace{7mm}

We wish to thank Wolfram Weise for making critical remarks on a preliminary
version of this manuscript. This research was partially supported by the
Israel Science Foundation grant No. 131/01.

\end{document}